\documentclass[letter,traditabstract]{aa}
\usepackage{graphicx}
\usepackage{rotating,subfigure,amssymb,afterpage}
\usepackage{txfonts}
\usepackage{natbib}
\usepackage[pdftitle={Local Underdensity in the{\sf REFLEX II} Survey},
pdfauthor={H. B\"ohringer et al.},
pdfsubject={}, pdfkeywords={}, bookmarks, bookmarksopen, 
colorlinks=true, linkcolor=blue, citecolor=blue, anchorcolor=blue,
urlcolor=blue]{hyperref} 
\newcommand{\propsim}{\lower 3pt \hbox{$\, \buildrel {\textstyle
      \propto}\over {\textstyle \sim}\,$}}
\usepackage{amstext}

\begin{document}

   \title{The extended ROSAT-ESO Flux-Limited X-ray Galaxy Cluster
   Survey (REFLEX II)\\ VI. Effect of massive neutrinos on the
cosmological constraints from clusters}

   \author{Hans B\"ohringer\inst{1}, Gayoung Chon\inst{1}} 

   \offprints{H. B\"ohringer, hxb@mpe.mpg.de}

   \institute{$^1$ Max-Planck-Institut f\"ur extraterrestrische Physik,
                   D-85748 Garching, Germany.
}

   \date{Submitted 29/11/14}

\abstract{Clusters of galaxies are important probes for the large-scale 
structure that allow us to test cosmological models. With the {\sf REFLEX II}
galaxy cluster survey we previously derived tight constraints
on the cosmological parameters for the matter density, $\Omega_m$,
and the amplitude parameter of the matter density fluctuations,
$\sigma_8$. Whereas in these previous studies no effect of massive neutrinos
was taken into account, we explore these effects in the present publication.
We derive cosmological constraints for the sum of the neutrino masses
of the conventional three neutrino families in the range 
$M_{\nu} =  \sum_i  m_{\nu~i} = 0$ to 0.6 eV. The influence on the
constraints of $\Omega_m$ and $\sigma_8$ for the expected mass range is
weak. Interesting constraints on the neutrino properties can be
derived by comparing the cluster data with those from
the {\sf Planck} cosmic microwave background (CMB) observations. The current
tension between the {\sf Planck} results and clusters can formally
be resolved with neutrino masses of about $M_{\nu} = 0.45 (\pm 0.28, 1\sigma)$ eV.
While we caution not to consider this a firm measurement because
it might also be the result of unresolved systematics,
it is interesting that other measurements of the local large-scale 
structure fluctuation amplitude, like that of cosmic lensing shear, yield 
similar results and additionally confirm the effect of massive neutrinos.
Among the indicators for massive neutrinos, galaxy clusters and in particular 
our large and well-controlled cluster survey currently provide the best
potential for constraints of the total neutrino mass.}


\authorrunning{B\"ohringer et al.}
\titlerunning{{\sf REFLEX II} cosmology constraints with massive neutrinos}
   \maketitle
%
%________________________________________________________________

\section{Introduction}

Clusters of galaxies form from weak density fluctuations in the early
Universe on comoving scales of several Mpc in a well-defined way.
They can thus be used to statistically assess the
large-scale  structure. By applying this as a diagnostics, one can
constrain cosmological models that describe our Universe. 
We have been using galaxy clusters from our statistically 
complete survey of galaxy clusters, the {\sf REFLEX} (ROSAT-ESO
Flux-Limited X-ray) Cluster Survey (B\"ohringer et al. 2001, 2004,
2013) to obtain constraints on cosmological model parameters
(B\"ohringer et al. 2014a). We derived in particular tight constraints
on the matter density parameter, $\Omega_m$, and on the 
amplitude parameter of the matter density fluctuations, $\sigma_8$
from the X-ray luminosity function measured in the nearby Universe.

In these calculations we assumed that neutrinos, filling the 
Universe in large abundance, have zero mass. Modern experimental
results from solar, atmospheric, and reactor neutrinos 
(e.g. Fogli et al. 2012; Forero et al. 2012; Lesgourgues et al. 2013)
show neutrino oscillations that imply that neutrinos are not 
massless; they therefore have an effect on the large-scale
structure of the Universe (e.g. Lesgourges \& Pastor 2006, 2014).
The process by which neutrinos influence the large-scale 
structure growth is a partial damping of the density fluctuation 
power spectrum on small and intermediate scales up to about
$k \ge 0.02~h$ Mpc$^{-1}$ ($\le 160~h^{-1}$ Mpc) in comoving 
units (Lesgourges \& Pastor 2006). 
Since the number of neutrinos in
the present Universe is approximately fixed, the strength of
the effect depends on the mass of the neutrinos. This mass is
unevenly distributed between the three different neutrino
species, but how exactly they share the mass is currently unknown.
For our purposes the important parameter is the total mass
of all neutrino families, $M_{\nu} =  \sum_i  m_{\nu~i}$,
and the distribution of the individual masses causes only
higher order effects that are of no concern for our calculations.
In addition, the effects of possible sterile neutrinos 
on the large-scale structure have been
considered (e.g. Lesgourges \& Pastor 2006, 2014).
In this paper we focus on the effect of three conventional
neutrinos.

There is an experimental lower limit on $M_{\nu}$
with a value of about 0.06 eV (e.g. Fogli et al. 2012; 
Forero et al. 2012). Several upper limits have 
been derived from astronomical observations with 
$M_{\nu} \le 0.93$ eV from {\sf Planck} alone (Planck Collaboration
XVI 2913a), $M_{\nu} \le 0.9$ eV from the SDSS power spectrum alone
(Viel et al. 2010), a value of $M_{\nu} = 0.36\pm 0.14~ (1\sigma)$ eV 
from the SDSS III power spectrum combined with {\sf WMAP} (Beutler et al. 2014),
$M_{\nu} \le 0.34$ eV  from SDSS III combined with CMB and supernova 
data (Zhao et al. 2014), $M_{\nu} \le 0.18$ eV from the matter power
spectrum of the WiggleZ Dark Energy Survey combined with data from
{\sf Planck} and other BAO observations (Riemer-S\o rensen 2014),
and $M_{\nu} \le 0.33$ eV from a combination of the cluster 
mass function, CMB, supernova, and BAO data (Mantz et al. 2010). 
One of the tightest constraints comes from the large-scale
structure analysis of the Ly$\alpha$ forest together with
galaxy clustering, CMB, and supernova observations with 
$M_{\nu}\le 0.17$ eV (Seljak et al. 2006). In a recent paper,
Costanzi et al. (2014) obtained constraints on a non-zero neutrino
mass of $M_{\nu} = 0.29 {+0.18 \atop -0.21} 
\left( 0.22 {+0.17 \atop -0.18}\right)$ eV using WMAP9 
({\sf Planck}) CMB as well as BAO, large-scale structure lensing shear
and cluster data. Hamann and Hasenkamp (2013) and
Battye and Moss (2014) also concluded on a positive signal for
neutrino mass from CMB and lensing shear or cluster data.

Therefore massive neutrinos
should be included in the modelling of the X-ray luminosity
function to comply with the most recent results, which imply
that neutrinos have mass. With a given neutrino number density we
can calculate the contribution of neutrinos to the matter density
in terms of the critical density and we find 
(e.g. Lesgourges \& Pastor 2014)
\begin{equation}
\Omega_{\nu} = { \sum m_{\nu}  \over 93.14 {\rm eV} }~  h^2~~ . 
\end{equation}
 
The aim of this paper is to explore how the constraints of 
cosmological parameters based on the {\sf REFLEX II} survey data
change by including massive neutrinos. While the effect of 
massive neutrinos on the cluster mass
function and on the derived cosmological constraints has
been explored before (e.g. Marulli et al. 2011,
Costanzi et al. 2013, Burenin 2013), the application
of the {\sf REFLEX II} cluster sample adds a new dimension to the
discussion for us. The {\sf REFLEX II} galaxy cluster sample
currently provides the most precise description of the shape of
the X-ray luminosity function. The better the shape of the function
is constrained, the better the degeneracy of the constraints on
the parameters, $\Omega_m$, $\sigma_8$, and $M_{\nu}$ 
can be broken. It is a main goal of this paper to explore 
what this implies for our data. In addition, we also wish
to study the possible implications on the neutrino mass
that can be gained by a comparison of the cosmological
constraints derived from our cluster sample with the 
results from the {\sf Planck} observations of the CMB.

The paper is structured as follows: In Sect. 2 we describe
the {\sf REFLEX II} galaxy cluster sample and in Sect. 3 the
cosmological modelling of these data.
In Sect. 4 we then discuss the effects of massive neutrinos 
on the X-ray luminosity
function, and in Sect. 5 we compare the cosmological constraints
from clusters including massive neutrinos with those from
the {\sf Planck} CMB observations and draw our conclusions. Section 6 provides a summary.
To determine all parameters that depend on distance, we use
a flat $\Lambda$CDM cosmology with a matter density parameter
$\Omega_m$ as required by the model. Literature values 
quoted above have  a scaling by $h = H_0/ 100$ km s$^{-1}$
Mpc$^{-1}$, whereas the following results are scaled with 
$h_{70} = h/0.7$, if not stated otherwise. For the comparison with 
{\sf Planck} we use a flat cosmology with $\Omega_m = 0.315$
and $h = 0.673$.

\section{REFLEX II galaxy cluster survey}

The {\sf REFLEX II} galaxy cluster survey is based on the X-ray detection of
galaxy clusters in the ROSAT All-Sky Survey (Tr\"umper 1993, 
Voges et al. 1999). The region of the survey is
the southern sky below equatorial latitude +2.5$^o$ and at Galactic
latitude $|b_{II}| \ge 20^o$. The regions of the Magellanic Clouds have
been excised. The survey region selection, the source detection, the
galaxy cluster sample definition and compilation, and the construction of
the survey selection function  as well as tests of the completeness of the
survey are described in B\"ohringer et al. (2013). In summary, the 
survey area is $ \sim 4.24$ ster. The nominal flux limit down to which
galaxy clusters have been identified in the ROSAT All-Sky Survey 
in this region is
$1.8 \times 10^{-12}$ erg s$^{-1}$ cm$^{-2}$ in the
0.1 - 2.4 keV energy band, yielding a catalogue of 911 clusters. 
To assess the large-scale structure
in this paper, we applied an additional cut
on the minimum number of detected source photons of 20 counts. This has
the effect that the nominal flux cut quoted above is only reached in about
80\% of the survey. In regions with lower exposure and higher interstellar
absorption, the flux limit is accordingly higher 
(see Fig.\ 11 in B\"ohringer et al. 2013). This effect is modelled and
taken into account in the survey selection function.

The flux limit imposed on the survey is for a nominal flux that has been
calculated from the detected photon count rate for a cluster X-ray spectrum
characterized by a temperature of 5 keV, a metallicity of 0.3 solar,
a redshift of zero, and an interstellar absorption column
density given by the 21cm sky survey described
by Dickey and Lockmann (1990). The result of this 
conversion of count rate to flux is an appropriate flux 
estimate before any redshift information and is analogous 
to an observed object magnitude corrected for Galactic extinction 
in the optical.

After the redshifts were measured, a new flux was calculated taking the
redshifted spectrum and an estimate for the spectral temperature
into account. The temperature was estimated by means of the X-ray luminosity -
temperature relation from Pratt et al. (2009) determined from the 
{\sf REXCESS} cluster sample. This is a sample of clusters 
drawn from {\sf REFLEX I} for deeper follow-up
observations with XMM-Newton, which is representative of the entire flux-limited
survey (B\"ohringer et al. 2007). The luminosity was determined first from the
observed flux by means of the luminosity distance for a given redshift. Using
the X-ray luminosity mass relation given in Pratt et al. (2009), we then
used the mass estimate to determine a fiducial radius of the cluster, which is
taken to be $r_{500}$ \footnote{$r_{500}$ is the radius where the average
mass density inside reaches a value of 500 times the critical density
of the Universe at the epoch of observation.}. We applied a beta model for the
cluster surface brightness distribution to correct for the possibly missing
flux in the region between the detection aperture of the source photons and
the radius $r_{500}$. The procedure to determine the flux, the luminosity,
the temperature estimate, and $r_{500}$ was performed iteratively and is described in
detail in B\"ohringer et al. (2013). In this paper we deduced a mean flux 
uncertainty for the {\sf REFLEX II} clusters of 20.6\%, which is 
mostly due to the Poisson statistics of the source counts, but also 
contains some systematic errors. 

The X-ray source detection and selection was based on the official
ROSAT All-Sky Survey source catalogue by Voges et al. (1999). 
We used the publicly available
final source catalogue \footnote{The RASS source catalogues can
be found at http://www.xray.mpe.mpg.de/rosat/survey/rass-bsc/  for the
bright sources and at http://www.xray.mpe.mpg.de/rosat/survey/rass-fsc/ for
the faint sources.} as well as a preliminary source list that
was created while producing the public catalogue.
To improve the quality of the source parameters for the mostly
extended cluster sources, we reanalysed all the X-ray sources with the
growth curve analysis method (B\"ohringer et al. 2000). The flux cut 
was imposed on the reanalysed data set. The process of the source
identification is described in detail in B\"ohringer et al. (2013). 

\section{Cosmological modelling of the REFLEX survey}

Cosmological constraints were obtained by comparing cosmological
model predictions for the galaxy cluster X-ray luminosity function
with the observations from the {\sf REFLEX II} project. 
The comparison was performed by means of a likelihood method.
The details of this procedure are described in 
B\"ohringer et al. (2014a), and we here provide only a brief outline. 
In a first step, the power spectrum of the matter density
fluctuations for the present epoch is calculated with the 
program CAMB (Lewis et al. 2000)
\footnote{ CAMB is publicly available from
http://www.camb.info/CAMBsubmit.html}. This is different to the 
previous calculations in B\"ohringer et al. (2014a), where we used 
the program for the transfer function by Eisenstein and Hu (1998).
By changing the calculations from the latter program to CAMB,
we did not note any differences larger than one percent. 
Based on this power spectrum, we calculated the cluster 
mass function with the formulas
given by Tinker et al. (2008). To derive the predicted X-ray 
luminosity function from the theoretically calculated cluster
mass function, we used empirical scaling relations of 
X-ray luminosity and mass in accordance with the observations
of Reiprich \& B\"ohringer (2002), Pratt et al. (2009), and
Vikhlinin et al. (2009), within their confidence limits.
In the marginalisations of the constraints we allowed for a
7\% uncertainty in the slope of the scaling relation and an
uncertainty of 14\% in its normalisation (equivalent to the mass 
calibration) as 1$\sigma$ constraints of these parameters.
For more details on the marginalisation see B\"ohringer et al. 
(2014a). In the final likelihood fit we do not compare 
the luminosity functions directly, but the comparison is 
made between the predicted and observed X-ray luminosity 
distribution. To theoretically predict the X-ray 
luminosity distribution, the X-ray luminosity function has 
to be folded with the survey selection function, which is also
described in detail in our previous paper.
  
Recent literature for example by Costanzi et al. (2014) suggested
that the galaxy cluster mass function should be modelled in
a particular way in the presence of massive neutrinos. The
suggested modification consists of only using the matter
density without neutrinos, $\rho_m - \rho_{\nu}$, in the 
relation of mass and filter radius to calculate 
the amplitude variance, $\sigma(M)^2$, of the density fluctuations.
We tested including this modification in our 
calculations and found that the results never changed by more 
than one percent. Since this is an order of magnitude smaller
than the systematic uncertainties, we did not include the 
modification at this stage.  

\section{Effect of neutrinos on the cluster X-ray 
luminosity function }

Before we describe the derived cosmological constraints,
we explore the effect of neutrinos on
the cluster abundance and the X-ray luminosity function 
of clusters. Neutrinos damp out large-scale structure
during the evolution of the Universe inside the horizon
scale as long as they are relativistic. Thus the greatest
length scale for which we expect damping effects is 
approximately the horizon scale at the epoch when the
neutrinos become non-relativistic. They can only damp
a fraction of the amplitude that corresponds to their fraction
of the total matter density. 

If we fix the normalisation of the power spectrum 
at the epoch of recombination with the parameter,  $A_S$,
the curvature power spectrum normalisation at a scale
of $k_0 = 0.05$ Mpc$^{-1}$, as done for the {\sf Planck}
power spectrum, we see that the present epoch power 
spectrum is depressed at small 
scales below a wave vector of about $0.014~h_{70} $ Mpc$^{-1}$
and the depression is stronger the larger $M_{\nu}$.
The regime that is relevant for cluster formation
is in the depression region. This is also the region 
in which the parameter $\sigma_8$ is determined. Therefore
we expect that the present-day value for $\sigma_8$
will change with changing neutrino masses for a fixed
$A_S$ normalisation. The ratio between these two parameters
also depends on the matter density, $\Omega_m$, since the 
maximum of the power spectrum shifts with this parameter.

\begin{figure}
   \includegraphics[width=\columnwidth]{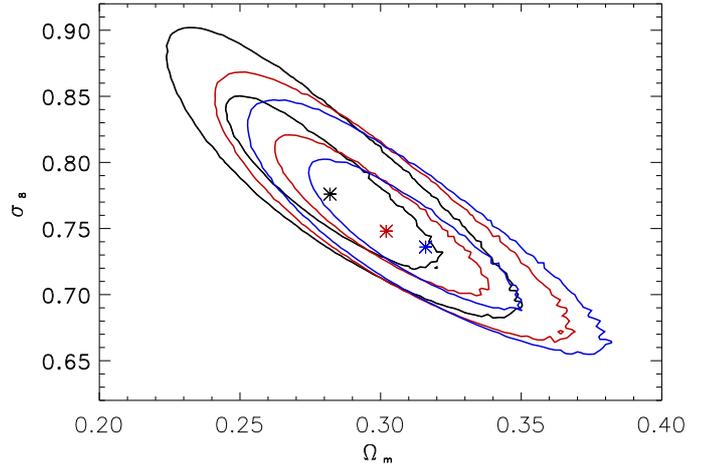}
\caption{Constraints on the cosmological parameters 
$\Omega_m$ and $\sigma_8$ from a model fit to the observed 
{\sf REFLEX II} X-ray luminosity distribution. The curves give
1 and 2$\sigma$ constraints for models with $M_{\nu} = 0$, 
$M_{\nu} = 0.4$, and $M_{\nu} = 0.6$ eV for the set of contours
from upper left to lower right (black, red, blue), respectively.
}\label{fig1}
\end{figure}

The parameter $\sigma_8$ was originally designed to describe
the power spectrum amplitude at cluster scale. Thus to first 
order, ignoring subtle changes in the shape of the 
renormalised power spectrum,
the value of $\sigma_8$ fixes the cluster abundance. 
Therefore looking for the best-fitting $\sigma_8$ for a given
cluster abundance means in the modelling that power spectra
for different neutrino masses will be 
rescaled such that they all 
give a very similar value of  $\sigma_8$. This is reflected
in the constraints we obtain for the parameter combination
of  $\Omega_m$ and $\sigma_8$ adopting different values of 
$M_{\nu}$ , as shown in Fig. 1. The subtle changes in the shape of
the power spectrum cause small moves in the $\Omega_m$ and $\sigma_8$
parameter plane, but the changes are much smaller than the changes
of $\sigma_8$ with neutrino mass for fixed $A_S$.
While the shift in $\sigma_8$  from $M_{\nu} = 0$ to
$M_{\nu} = 0.6$ eV in Fig. 1 is $\Delta \sigma_8 = 0.04$, it is about
$\Delta \sigma_8 = 0.14$ for fixed $A_S$. For the results in Fig. 1
we have considered extreme cases. If the possible range of total 
neutrino masses is instead about 0.06 to 0.2 eV, we expect 
differences in the marginalisation results smaller than the 
present marginalisation uncertainties, if we use the 
$\sigma_8$-normalisation of the power spectrum.

There is also hardly any distinction in the goodness of fit
between the fits for different $M_{\nu}$. The 
likelihood changes by less than $\Delta L=1,$ which is well
within the one-sigma errors. Taking the cluster results 
alone therefore does not yield a clear preference for a 
neutrino mass in the mass range shown. A discrimination can 
be obtained by comparing this with the observations of the CMB, 
which will be discussed in the next section.

\section{Comparison with the Planck CMB results}

\begin{figure}
   \includegraphics[width=\columnwidth]{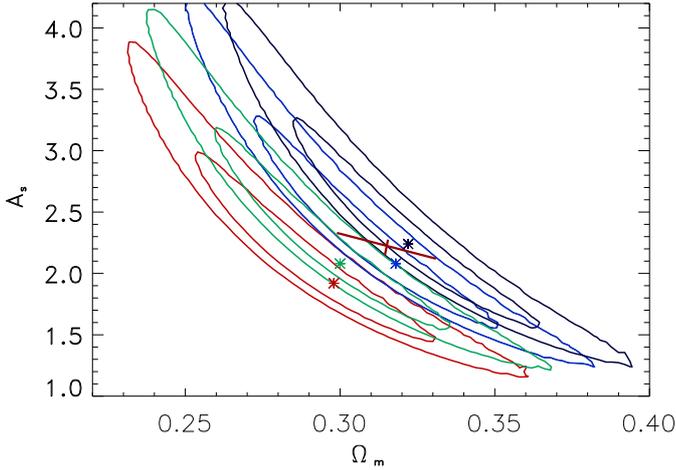}
\caption{Constraints on $A_s$ and $\sigma_8$
from the {\sf REFLEX II} X-ray luminosity function for values
of $M_{\nu} = 0$, 0.17, 0.4, and 0.6 eV for the contours
from bottom to top, respectively. The contours give the 1 and
$2\sigma$ confidence intervals. We also show the
constraints derived from the {\sf Planck} CMB observations
(data point with $1\sigma$ error bars).
}\label{fig2}
\end{figure}

To compare the present constraints with the {\sf Planck} results,
we chose a power spectrum normalisation
that is applicable in the same way to both surveys. For
this reason, we used the parameter, $A_S$, the normalisation of
the dimensionless curvature power spectrum, to normalise the 
power spectrum for the cluster abundance calculation, since
this parameter is also used to normalise the power spectrum 
for the {\sf Planck} data analysis (Planck Collaboration
2013a). With this parameter we calculated the matter power
spectrum for the present epoch, folded it through the
structure formation modelling and fitted it to the observed
{\sf REFLEX} X-ray luminosity function. In this way, we
derived constraints on cosmological parameters in the
$\Omega_m$ - $A_S$ plane as displayed in Fig. 2.
The plot shows the marginalisation results for 
four different total neutrino masses, $M_{\nu} = 0, 0.17, 0.4,  \text{and } 
0.6$ eV. As for the $\Omega_m$ - $\sigma_8$ 
constraints, the two parameters are somewhat degenerate. Similar to the data in Fig. 1, the likelihoods of
the minima for the different contour sets feature a 
difference lower than $\Delta L \le 1$. Thus within 1$\sigma$
uncertainty limits, we cannot distinguish the goodness 
of fit between the different models. Taking the cluster 
data alone, all these neutrino masses are possible.

In Fig. 2 we compare the cluster 
constraints with the constraints from the analysis of the
power spectrum of CMB anisotropies seen by {\sf Planck}
(Planck Collaboration 2013a). The {\sf Planck} data point 
that represents the results for {\sf Planck}+WP
in Table 2 of Planck Collaboration (2013a) 
is shown with $1\sigma$ error bars. The 
{\sf Planck} results have been derived for a cosmology with
$M_{\nu} = 0.06$ eV, while the value for $M_{\nu}$ was varied
for the cluster constraints. We checked that the CMB
power spectrum does not vary in any significant way with
a variation of $M_{\nu}$ in the considered range for fixed $A_s$. 
Therefore the cluster constraints for different values of
$M_{\nu}$ can be compared with one representation of the 
{\sf Planck} data in this plot. The tilt of
the cross of error bars for {\sf Planck} follows the 
shape of the error ellipse.
Its orientation is taken from the error ellipse shown in Fig. 11
of the publication by the Planck Collaboration (2013b) 
and converted from the $\Omega_m - \sigma_8$ to the 
$\Omega_m$ - $A_S$ representation.

Formally, the two data sets
can be reconciled for a total neutrino mass in the range 
$M_{\nu} = 0.45 (\pm  0.28)$ eV including the combined 
1$\sigma$ uncertainties of both data sets (in a 
conservative way with a direct addition of the errors instead 
of a Gaussian addition). 
\footnote{The numerical values were obtained from the 
relative location of the error ellipses determined on a finer
grid of $M_{\nu}$ values, than as shown in Fig. 2. For this 
evaluation the exact orientation of the {\sf Planck} error
ellipse is crucial.}
A similar conclusion was reached
by the Planck Collaboration (2013b): cluster and CMB
data can be reconciled with a neutrino mass of about 
$M_{\nu} = 0.58 (\pm  0.2)$. We caution,
however, not to interpret this result too quickly as a constraint on the
neutrino masses because it might in principle also be
the result of systematic uncertainties and calibration problems.
But the results definitely illustrate the power
of combining the two cosmological probes to determine the neutrino masses.

\begin{figure}
   \includegraphics[width=\columnwidth]{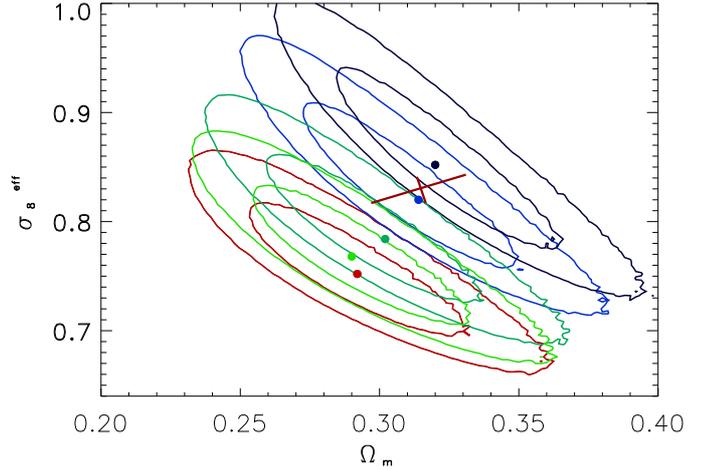}
\caption{Marginalised constraints for the rescaled
 parameter $\sigma_{8}^{eff}$ and $\Omega_m$ for values
of $M_{\nu} = 0$, 0.06, 0.17, 0.4, and 0.6 eV for the contours
from bottom to top, respectively. The contours give the 1 and
$2\sigma$ confidence intervals. We also show the result of the 
{\sf Planck} CMB observations
as data point with $1\sigma$ error bars.
}\label{fig3}
\end{figure}

Most illustrations in the literature show constraints
for the $\sigma_8 - \Omega_m$ diagram. This was also
made for the cluster analysis of the clusters detected through
the Sunyaev-Zeldovich effect with {\sf Planck} in 
comparison with the {\sf Planck} CMB results (Planck Collaboration
2013b). Therefore we tried to find a way to translate the data shown
in Fig. 2 into a plot of $\sigma_8 - \Omega_m$ constraints.
This can be achieved by translating the $\Omega_m$ - $A_S$
results into a $\sigma_8 - \Omega_m$ constraint in a representation
that keeps the value of $M_{\nu}$ fixed to 0.06 eV. This is 
analogous to an analysis of cosmological data as a function
of the Hubble parameter, but choosing a representation
in which the results are translated into a cosmology with
a fixed value for $H_0$.
We achieved this by using for $\sigma_8$
not the value we would measure at present for the given 
cosmology and $M_{\nu}$ value, but instead we used a new
parameter, $\sigma_{8}^{eff}$, the $\sigma_8$ value that would be
predicted on the bases of the $A_s$ normalisation and the
cosmological model used for the best fit of the {\sf Planck}
results with $h = 0.673$, $\Omega_m = 0.315$ and $M_{\nu}=0.06$ eV.
The new parameter is given by

\begin{equation}
\sigma_{8}^{eff} = \sigma_{8}^{true} \times {\sigma_{8}^{ref} \over \sigma_{8}^{mod}} 
,\end{equation}

where $\sigma_{8}^{ref}$ is the value from the {\sf Planck} cosmology reference
model for fixed $A_s$ and $\sigma_{8}^{mod}$ the value for a similar
model with the correct value for $M_{\nu}$ and the same $A_s$. 
The resulting plot
is shown in Fig. 3. The position of the {\sf Planck} data
point with respect to the cluster constraint contours is 
equivalent to the situation in Fig. 2, and the orientation
of the error ellipse has been taken into account. The plot
shows that the same relative location of the error contours and thus
the conclusions on the constraints on $M_{\nu}$ are the same as gained from Fig. 2.

\section{Summary and conclusion}

Based on the {\sf REFLEX II} galaxy cluster survey and its well-defined X-ray luminosity function, we derived tight 
constraints on the cosmological parameters  $\sigma_{8}$ and $\Omega_m$ 
and studied the influence of massive neutrinos on these results.
Within the limits of the expected range of the total mass of the
conventional three neutrino families of about $M_{\nu} = 0.06 - 0.2$ eV,
we found only weak changes of $\sigma_{8}$ and $\Omega_m$ with
neutrino mass. The changes are within the limits of the current
uncertainties, and there is no preference for a certain total neutrino
mass from the galaxy clusters alone.

The constraints become more interesting when the cluster data
are combined with observations of the CMB anisotropies by {\sf Planck}.
For this comparison we performed the cosmological parameter
constraints for the parameter combination $A_s$ and $\Omega_m$.
Without massive neutrinos there is a discrepancy between  
the results from the two data sets, as discussed previously
(Planck Collaboration 2013b; B\"ohringer 2014a). When massive
neutrinos are included, the discrepancy can formally be reconciled for a total
neutrino mass of $M_{\nu} = 0.45 (\pm  0.28)$ eV. It is interesting
that a discrepancy between the CMB results and other measurements
of the present-day large-scale structure amplitude have been 
found within the $\Lambda$CDM model without massive neutrinos.
For example, Battye and Moss (2014) found constraints on the total 
neutrino mass of $M_{\nu} = 0.320 (\pm 0.081)$ eV for the combination 
of {\sf Planck} CMB data and lensing shear from the CFHTLens survey
that agree well with our findings.
Hamann and Hasenkamp (2013) found a similar tension in a massless
neutrino cosmology for the combination of CMB and clusters as well as  
CMB and cosmic shear. In their analysis, they only investigated sterile neutrinos because these simultaneously decrease 
the tension in the results for the Hubble constant, and thus their
result is not directly comparable with ours. Both Hamann and Hasenkamp
and Costanzi et al. (2014) pointed out that the strongest
driver for a positive neutrino mass comes from clusters. Since 
our results on the cluster abundance are among the most precise 
results, they will contribute to the strongest constraints for
the total neutrino mass.

We here only considered classical 
neutrinos. The main driver to include sterile neutrinos in recent
publications (e.g. Hamann and Hasenkamp 2013) is the difference
on the Hubble parameter measured by {\sf Planck} and locally
with calibrated distance indicators such as the Cepheides. We have
shown in a recent paper using our cluster data that there
are indications that we live in a locally underdense region
of the Universe in which one expects the Hubble parameter 
to be locally higher (B\"ohringer et al. 2014b). It is worth
noting that this can resolve some of the tension between the
local and global measurement of $H_0$. This also makes the
results with massive non-sterile neutrinos more attractive. 

The present marginalised constraints from the galaxy cluster data
take the uncertainties in the scaling relations, the most serious 
bottle-neck preventing us from deriving tighter constraints,
into account in a fairly conservative way (B\"ohringer et al. 2014a).
Much effort is currently made to better constrain the 
X-ray scaling relations of galaxy clusters with deeper observations
of well-selected cluster samples with XMM-Newton and Chandra
and to improve the cluster mass calibration with weak-lensing studies.
For the {\sf Planck} data intense calibration efforts are ongoing
as well.
We therefore expect a significant improvement of the understanding of
the systematics in the near future, which will allow us to exploit
the full potential of the observational data.

\begin{acknowledgements}
H.B. and G.C. acknowledge support from the DFG Transregio Program TR33
and the Munich Excellence Cluster ''Structure and Evolution of the Universe''.  
G.C. acknowledges support by the DLR under grant no. 50 OR 1405.
\end{acknowledgements}

\end{document}